\documentclass[12]{article}
\usepackage{dcolumn}
\usepackage{color}
\usepackage{latexsym}
\usepackage{bm}
\begin{document}
\date{}
\title{\textbf{Hawking Fluxes, Back reaction and Covariant 
 Anomalies}}
\author{{Shailesh Kulkarni}\thanks{E-mail:shailesh@bose.res.in}\\
\textit{S.~N.~Bose National Centre for Basic Sciences,}\\
\textit{JD Block, Sector III, Salt Lake, Kolkata-700098, India}}
\maketitle
\begin{quotation}
\noindent \normalsize
 Starting from the chiral covariant effective action approach
 of Banerjee and Kulkarni [Phys. Lett. B 659, 827(2008)],
 we provide a derivation of the Hawking radiation from 
 a charged black hole in the presence of gravitational back reaction. The modified expressions for
 charge and energy flux, due to effect of one loop back reaction are obtained. 
\end{quotation}
{\it Introduction :}\\

Hawking radiation is an important and prominent quantum effect that arises upon the quantization of matter fields in a background spacetime with an event horizon. 
This is due to the different vacuum states between the vicinity of the event horizon and 
 the asymptotic infinity \cite{Hawking}. Apart from the
 original derivation \cite{Hawking}, there are other approaches \cite{gibbons,parikh} each having their own merits and demerits. Recently, Robinson and Wilczek
\cite{Robwilczek} followed by Iso, Umetsu and Wilczek \cite{Isowilczek}, gave a new derivation of Hawking effect based on anomaly cancellation approach. Soon after, the 
  analysis of \cite{Robwilczek,Isowilczek} was reformulated completely by Banerjee and Kulkarni \cite{shailesh} in terms of covariant expressions only. Very recently,
 Banerjee and Kulkarni \cite{shailesh2, rabin} gave a new derivation of the Hawking effect based solely on the structure of the effective action and boundary conditions at event horizon. 
 None of these computations however, consider the effect
 of back reaction.

         Back reaction, it might be recalled is an effect of non-zero expectation value of the 
 energy-momentum tensor on the spacetime geometry, which acts as a source of curvature. 
  It is possible to include the effect of gravitational back reaction in the derivation of Hawking radiation. Indeed, using the conformal anomaly method
 the effect on  the spacetime geometry by one loop back reaction was computed in \cite{york, lousto}. Based on this approach, corrections to the Hawking temperature
 were obtained in \cite{lousto,fursaev}. Correction to the Hawking temperature using the back reaction equation for linearised quantum fluctuation was derived in \cite{fabbri}.
 Recently, more useful and intuitive way to understand the effect of back reaction through the  quantum tunnelling formalism \cite{parikh} was developed in \cite{bibhas}. Naturally,
 it becomes interesting to incorporate the effect of back reaction in the analysis of \cite{Robwilczek,shailesh2,shailesh}.
     
      A particularly useful way to understand the Hawking effect is through effective action formalism developed in \cite{shailesh}. Only the structure of effective action defined near the event horizon is sufficient to determine the Hawking flux. An important ingredient of 
 \cite{shailesh}  was to realize that effective theory become two dimensional  and chiral near the event horizon \cite{Robwilczek}. 
 Yet another important aspect in this approach was that,
 implementation of a specific boundary condition - the vanishing of the covariant form of the current and energy momentum tensor \cite{shailesh}. Unlike the approaches \cite{Robwilczek,Isowilczek,shailesh}, the important advantage of 
 this method\cite{shailesh2} was that, Hawking fluxes were obtained only
 by using the covariant anomaly near the event horizon. 
This was true not just for charged black holes, but for other spacetime geometries as well \cite{sunandan,shailesh3,peng} and 
 also for deriving the Hawking fluxes associated with higher spin\cite{Iso}.
        
             In this paper we present a derivation for Hawking energy and charge flux for the Reissner-Nordstrom black hole in the presence of back reaction, by using
 the effective action approach developed in \cite{shailesh2}. The expressions for charge and energy flux get modified as a consequence of the one loop effect of back reaction.\\ 
 
{\it Back reaction and Hawking fluxes :}\\
 
 We are interested in discussing the Hawking effect from a charged black hole in the presence of back reaction. The metric for charged black hole is given by
\begin{equation}
ds^2_{o} = f(r)dt^2 - \frac{1}{f(r)}dr^2 - r^2 d\Omega^2, \label{0.1} 
\end{equation}               
where the function $f(r) = \frac{(r-r_{H})(r-r_{i})}{r^2}$ admits inner$(r=r_{i})$ and outer$(r=r_{H})$ horizon $f(r_{i}) = f(r_{H}) = 0$ respectively and $d\Omega^2$
is the line element on $2$ sphere. The gauge potential is defined by $A = -\frac{Q}{r}dt$.\\

         We now consider, modification in the spacetime metric, from its usual form (\ref{0.1}), due to the one loop back reaction \cite{lousto}. We consider the 
 most general, static  spherically symmetric metric 
\begin{equation}
ds^2 = B(r)dt^2 - \frac{1}{H(r)}dr^2 - r^{2}d\Omega^2, \label{0.2}  
\end{equation}      
where $H(r)$ and $B(r)$ are the metric coefficients\footnote{An explicit form for  the metric is given in \cite{lousto}. In our analysis, however, we use only the general properties of metric coefficients like $B(r_{\infty})H(r_{\infty}) =1$. The explicit structure of the metric is not crucial for our
 purpose.} on the $r-t$ sector and $d\Omega^2$ is the line element on the 2-sphere. The horizon is now defined by $B(r=r_{+}) = H(r=r_{+}) = 0$ and $r_{+}$ is the modified horizon radius given by \cite{lousto} 
\begin{equation}
r_{+} = r_{H}\left(1 + \frac{\beta}{M^2}\right) \label{0.21}
\end{equation}
where $r_{H} = M + \sqrt{M^2 -Q^2}$ is the usual (outer) horizon radius  for the charged black hole metric (\ref{0.1}). Such a form is dictated
by simple scaling arguments. As is well known, a loop expansion is equivalent to an expansion
in powers of the Planck constant $h$. Since, in natural units, $\sqrt{h} = Mp$, the one loop correction
has a form given by $\frac{\beta}{M^2}$ . The parameter $\beta$ is the 
 negative constant related to the trace anomaly coefficient taking into
account the degrees of freedom of the fields and its explicit form is 
given in \cite{york,lousto}. Note that the modified horizon radius gets decreased
 with respect to its usual value. Here we would like to mention that the generic 
 form for the metric (\ref{0.2}) in the presence of the gravitational back reaction was obtained \cite{york, lousto}
 by solving the semiclassical Einstein equations
\begin{equation}
R_{\mu\nu} -\frac{1}{2}g_{\mu\nu}R = \langle T_{\mu\nu}(\phi,g_{\mu\nu})\rangle \nonumber
\end{equation}
or in the more convenient form 
\begin{equation}
R_{\mu\nu} = \langle T_{\mu\nu}(\phi,g_{\mu\nu})\rangle - \frac{1}{2}\langle T^{\rho}_{\rho} \rangle g_{\mu\nu}, \label{einstein}
\end{equation}
with the aid of conformal (trace) anomaly in $4d$ and by keeping the spherical 
 symmetry intact.  
In our formalism, we consider the generic form for the $4d$ metric 
(\ref{0.2}), in the presence of the back reaction, as a starting point.
Since the anomaly can be shifted from conformal to the diffeomorphism anomaly,
it is expected that a similar form of the metric is obtained from the 
covariant anomaly.\\
      As mentioned before by using a dimensional reduction technique \cite{Robwilczek}, the effective field theory near the horizon  becomes a two dimensional chiral theory. The metric of this two dimensional theory is identical to the $r-t$ sector of the full metric (\ref{0.2}). Henceforth we shall always use $g_{\mu\nu}$ for the $r-t$ part
 of the metric (\ref{0.2}). Note that unlike the case of \cite{shailesh2},  we have $\sqrt{-g} \ne 1$ where $g = det{g_{\mu\nu}}$. Next, we consider anomalous (chiral) effective action \cite{Leut}, which describes the theory near the horizon. We then consider expressions for the covariant currents and  energy momentum tensor\cite{shailesh2}, obtained by taking appropriate functional derivative of the effective action. Unknown constants appearing in these solutions
 are fixed by employing the covariant boundary condition \cite{shailesh,shailesh2}. Once these constants are fixed, the charge and energy flux are obtained from the asymptotic limit of the current and energy momentum tensor respectively.\\

     First, we concentrate on the gauge sector. The expression for covariant current obtained from the anomalous
 (chiral) effective action is given by \cite{shailesh2,Leut}
\begin{equation}
J^{\mu}_{(H)} = -\frac{e^2}{2\pi} D^{\mu}\bar B \label{3.1} 
\end{equation}     
where $D_{\mu} = \nabla_{\mu} - \bar\epsilon_{\mu\rho}\nabla^{\rho}$ is a chiral derivative and $\bar B(x)$ is defined as
\begin{equation}
\bar B(x) = \int d^{2}y \sqrt{-g}\Delta^{-1}_{g}(x,y) \bar \epsilon^{\mu\nu}\partial_{\mu}A_{\nu}(y). \label{3.2}
\end{equation}
Note that by taking covariant divergence of (\ref{3.1}) we get the anomalous gauge ward identity  
\begin{equation}
\nabla_{\mu} J^{\mu} = -\frac{e^2}{2\pi \sqrt{-g}}\epsilon^{\rho\sigma}\partial_{\rho}A_{\sigma}. \label{3.21}
\end{equation}
The anomalous term on right side of (\ref{3.1}) is the covariant gauge anomaly \cite{bertlmann,Fujikawa,Bardeen}. Note further that $\bar B(x)$ defined in (\ref{3.2}) satisfy 
\begin{equation}
\Delta_{g}\bar B(x)  = \frac{\epsilon^{\mu\nu}}{\sqrt{-g}}\partial_{\mu}A_{\nu}(x). \label{3.3}
\end{equation}
where $\Delta_{g} = \nabla^{\mu}\nabla_{\mu}$ is the Laplacian for the $r-t$ sector of the metric (\ref{0.2}). 
 After explicitly calculating $\Delta_{g}$ and using the form for gauge potential, we arrive at
\begin{equation}
\partial_{\mu}(\sqrt{-g}g^{\mu\nu}\partial_{\nu}\bar B(x)) = - \partial_{r}A_{t}. \label{3.4}
\end{equation}
The solution of above differential equation is given by
\begin{equation}
\bar B = \bar B_{o}(r) - at + b \ ; \ \partial_{r}\bar B_{o} = \frac{1}{\sqrt{HB}}(A_{t}(r) + c)  \label{3.5}
\end{equation} 
where $a,b$ and $c$ are integration constants. Once we have the solution for $B$ we could easily obtain
 expression for the covariant current. Substituting (\ref{3.5}) in (\ref{3.1}) for $\mu=r$, yields
\begin{equation}
J^{r}_{(H)} = \frac{e^2}{2\pi}\sqrt{\frac{H}{B}}(A_{t}(r)+ c + a). \label{3.6}
\end{equation} 
Now we fix the value of $c + a$ by imposing the covariant boundary condition, namely, the covariant current vanishes at the event horizon defined by $r=r_{+}$. Thus by setting $J^{r}_{(H)}(r_{+}) = 0$ we get
\begin{equation}
c + a = - A_{t}(r_{+}). 
\end{equation}     
Hence, the expression for $J^{r}_{(H)}(r)$ becomes
\begin{equation}
J^{r}_{(H)}(r) = \frac{e^2}{2\pi}\sqrt{\frac{H}{B}}(A_{t}(r)
 - A_{t}(r_{+})). \label{3.7}
\end{equation} 
 Now the charge flux is determined by the asymptotic limit of the anomaly free current \cite{Isowilczek, shailesh}. As it is evident from the expression (\ref{3.21}) the anomaly vanishes in this limit and therefore we obtain the charge flux directly form (\ref{3.7}) by taking its asymptotic limit. This gives us
\begin{equation}
C_{o} = J^{r}_{(H)}(r \rightarrow \infty) = - \frac{e^2}{2\pi}
A_{t}(r_{+}). \label{3.8}
\end{equation}
Finally, using the form for gauge potential and substituting (\ref{0.21}) in (\ref{3.8}), yields
\begin{equation}
C_{o} = \frac{e^2 Q}{2\pi r_{H}(1 + \frac{\beta}{M^2})}. \label{3.9}
\end{equation}
This is the expression for Hawking charge flux. Further, expanding $(1 + \frac{\beta}{M^2})^{-1}$ and keeping only leading order term in $\beta$, we get 
\begin{equation}
C_{o} \approx \frac{e^2 Q}{2\pi r_{H}} -  \frac{e^2 Q\beta}{2\pi r_{H}M^2}. \label{3.10} 
\end{equation}
The first term in the above expression is the usual charge flux for charged black hole \cite{Isowilczek, shailesh, shailesh2} while the next term represent correction by the
 effect of one loop back reaction.\\

     Now we will concentrate our attention on the gravity sector. The expression for covariant energy momentum tensor obtained by considering the functional variation of the effective action \cite{shailesh2, Leut} is
\begin{eqnarray}
T^{\mu}_{\nu} = \frac{e^2}{4\pi}\left(D^{\mu} \bar B D_{\nu} \bar B\right) \nonumber\\
 +\frac{1}{4\pi}\left(\frac{1}{48}D^{\mu}G D_{\nu}G 
-\frac{1}{24} D^{\mu} D_{\nu}G + \frac{1}{24}\delta^{\mu}_{\nu}R\right)
\label{3.9} 
\end{eqnarray}  
where $\bar B$ is given by (\ref{3.2}) while $G(x)$ is defined as
 \begin{equation}
G(x) = \int d^2 y \ \sqrt{-g} \Delta_{g}^{-1}(x,y) R(y), \label{3.91}
\end{equation}
 and  
\begin{equation}
R = \frac{B''H}{B} + \frac{B'H'}{2B} - \frac{B'^2 H}{2B^2}. \label{3.92} 
\end{equation}
is the scalar curvature for the metric $g_{\mu\nu}$.

 Now by taking the covariant divergence of (\ref{3.9}) we get the anomalous gravitational Ward identity,
\begin{equation}
\nabla_{\mu}T^{\mu}_\nu = F_{\mu\nu}J^{\mu} + \frac{1}{96\pi} \epsilon_{\nu\mu}\partial^{\mu}R.\label{3.10}
\end{equation}
The first term on right hand side of above equation is classical Lorentz force while the second term is the covariant gravitational anomaly \cite{kohlprath}. Also, note that, by acting $\Delta_{g}$ on $G(x)$ defined in (\ref{3.91}), we get
\begin{equation}
\partial_{\mu}(\sqrt{-g}g^{\mu\nu}\partial_{\nu}G) = \sqrt{-g}R. \label{3.11}
\end{equation} 
The solution for the above equation is given as 
\begin{equation}
G= G_{o}(r) - 4pt + q  \ ; \ \partial_{r}G_{o} = 
\frac{-1}{\sqrt{HB}}\left(\sqrt{\frac{H}{B}}B' + z\right) \label{3.12} 
\end{equation}
where $p,q$ and $z$ are constants.

After using solutions for $B(x)$ and $G(x)$, the $r-t$ component of the covariant energy momentum tensor (\ref{3.9}) becomes,
\begin{eqnarray}
T^{r}_{t} &=& \frac{e^2}{4\pi\sqrt{-g}}(A_{t}(r) - A_{t}(r_{+}))^2 + \frac{1}{12\pi\sqrt{-g}}\left(p - \frac{1}{4}\left(\sqrt{\frac{H}{B}}B' + z\right)\right)^2\nonumber\\ 
&& + \frac{1}{24\pi\sqrt{-g}}\left[\sqrt{\frac{H}{B}}B'\left(p - \frac{1}{4}\left(\sqrt{\frac{H}{B}}B' + z\right)\right) + \frac{1}{4}HB'' - \frac{B'}{8}(\frac{HB'}{B} - H')\right].
\label{3.14} 
\end{eqnarray}
Now we implement the boundary condition, namely the vanishing of the covariant energy momentum tensor 
at the horizon, this condition leads to a relation among the unknown constants $p$ and $z$
\begin{equation}
p=\frac{1}{4}(z \pm \sqrt{H'_{+}B'_{+}}) \ ; \  H'_{+} \equiv H'(r = r_{+}) \ ; \ B'_{+}\equiv B'(r=r_{+}). \label{3.15} 
\end{equation}
Substituting either of the above solutions in (\ref{3.14}) we get 
\begin{eqnarray}
T^{r}_{t} &=& \frac{e^2}{4\pi\sqrt{-g}}(A_{t}(r) - A_{t}(r_{+}))^2 \nonumber\\
&& \frac{1}{192\pi\sqrt{-g}}\left[B'_{+}H'_{+} - \frac{2HB'^2}{B} + 2HB'' + B'H'\right] \label{3.16}  
\end{eqnarray}
Now the energy flux might be recall is given by the asymptotic expression for the anomaly free energy momentum
 tensor. As happened for the charge case here also we see from (\ref{3.10}) that the anomaly vanishes 
in $r\rightarrow\infty$ limit. Therefore the energy flux is obtained by taking asymptotic limit
 of the above equation. Then we get
\begin{equation}
a_{o} = T^{r}_{t}(r\rightarrow\infty)=\frac{e^2}{4\pi}A_{t}(r_{+})^2 +\frac{1}{192\pi}B'_{+}H'_{+}.\label{3.17}
\end{equation}
We would like point out that the expression for the energy-momentum
tensor derived from the  usual effective action induced by the conformal anomaly
in $4d$ is present in the literature \cite{fabbri1,fabbri2,mottola}
There the ingoing and the outgoing fluxes were obtained respectively
 by taking the near horizon and asymptotic limit of the energy-momentum tensor. On the other hand, in our approach, the Hawking fluxes are 
 obtained by taking the asymptotic limit of the anomalous energy-momentum tensor
derived from the chiral effective action in $2d$ and by implementing 
 the covariant boundary condition at the horizon. It is therefore difficult
to compare the $4d$ results with the one obtained in present approach.
However, our results are consistent with that obtained by taking the 
 expectation value of the usual (anomaly free) energy-momentum tensor derived from $2d$
 Polyakov effective action, in the Unruh vacuum \cite{fabbri1}
 \footnote{The connection between the covariant anomaly
 method and the approach based on the usual (Polyakov) effective action
 is discussed in \cite{shailesh2}.}. \\
       
Next, we can write the above expression in terms of the modified surface gravity \cite{lousto} as
\begin{equation}
a_{o} = \frac{e^2 Q^2}{4\pi r_{H}^2 (1 + \frac{\beta}{M^2})^2} + \frac{1}{48\pi}
K_{M}^2. \label{3.18} 
\end{equation}
where $K_{M} = \frac{1}{2}\sqrt{B'_{+}H'_{+}}$. Now following the similar
 argument below (\ref{0.21}), one loop correction to the surface gravity due
 to effect of self-gravitation, is given by
\begin{equation}
K_{M} = K_{o}\left(1 + \frac{\alpha}{M^2}\right), \label{3.19} 
\end{equation}
 where the parameter $\alpha$ like $\beta$ is related to the trace anomaly coefficient, given by \cite{fursaev}
\begin{equation}
\alpha = \frac{1}{360\pi}(-N_{o} -\frac{7}{4}N_{\frac{1}{2}} + 13N_{1} + \frac{233}{4}N_{\frac{3}{2}} - 212N_{2}) \label{alpha} 
\end{equation}
where $N_{s}$ denotes number of fields with spin $s$.  $K_{o}$ is the usual surface gravity (in the absence of the back reaction) for the charged
black hole. 

        Now by using  (\ref{3.18}) and (\ref{3.19}) we get
 \begin{equation}
a_{o} = \frac{e^2 Q^2}{4\pi r_{H}^2 (1 + \frac{\beta}{M^2})^2} + \frac{1}{48\pi}
K_{o}^{2}\left(1 + \frac{\alpha}{M^2}\right)^{2}
 = \frac{e^2 Q^2}{4\pi r_{H}^2 (1 + \frac{\beta}{M^2})^2}+\frac{\pi T_{o}^2}{12}(1+\frac{\alpha}{M^2})^2, \label{3.20} 
\end{equation}
where $T_{o} = \frac{K_{0}}{2\pi}$ is the usual Hawking temperature for Reissner-
Nordstrom black hole. The above equation represents the expression for the modified energy flux by the effect of one loop back reaction. Further, we can recast the 
 above equation by expanding upto leading order in $\alpha$
 and $\beta$. This yields 
\begin{equation}
a_{o} \approx \frac{e^2 Q^2}{4\pi r_{H}^2} + \frac{\pi T_{o}^2}{12}
 - \frac{e^2 Q^2 \beta}{2\pi r_{H}^2 M^2} + \frac{\pi T_{o}^2 \alpha}{6M^2}. \label{3.21}  
\end{equation}
The first two terms in the above expression represent
 energy flux from the usual charged black hole \cite{Isowilczek, shailesh2, shailesh}, while the last two terms  are corrections due to the effect of one loop back reaction. Also, by using (\ref{3.21}) and noting that $\beta$ is negative, we observe that, if $\alpha>0$, the Hawking flux is increased by the back reaction from its standard value. While, on the other hand
 if $\alpha<0$ i.e when the graviton contribution is dominant, a different 
scenario will appear here. Indeed, for $\alpha<0$ we can write the last term
 of (\ref{3.20}) as $\frac{\pi}{12}T_{o}^2\left(1-\frac{|\alpha|}{M^2}\right)^2$. Therefore, when the graviton contribution is dominant there is net 
 decrease in the Hawking flux, due to one loop back reaction effect,
 from its usual value. \\  
{\it Discussions :}\\
 Based on the effective action approach we have given a derivation of the Hawking charge and energy flux
 for the Reissner-Nordstrom black hole taking into the account the effect of one loop back reaction. 
 The point is that the $r-t$ part of usual charged black hole ($\sqrt{-g} =1$) gets modified to a more general ($\sqrt{-g}\ne 1$) due to the effect of back reaction without disturbing the spherical symmetry. For this general metric, the 
   expressions for the covariant current and energy momentum tensor were obtained. This indicates the generality of
 the effective action approach developed in \cite{shailesh2}. The corrections to charge and energy flux due to
 (one loop) back reaction effect were then obtained by appropriately taking  asymptotic limit of the current and energy momentum tensor. The enhancement 
 in the Hawking temperature due to the effect of back reaction was reported 
 in \cite{lousto}. Here, we note that, since $\beta<0$, the Hawking flux (\ref{3.18}), for $\alpha>0$ (i.e when matter is dominant) gets enhanced by the effect of back reaction, while for $\alpha<0$ it gets reduced by the 
 one loop back reaction effect. Consequently, the mass-loss and the life time of the  Reissner-Nordstrom black hole would be modified. A similar reasoning holds for  the charge flux obtained in (\ref{3.10}). 
\\

{\it Acknowledgement:} I wish to thank Prof. R. Banerjee
for useful suggestions and constant encouragement throughout this work.



\begin{thebibliography}{99} 
\bibitem{Hawking} S. Hawking, Commun. Math. Phys. {\bf 43}, 199 (1975)
\bibitem{gibbons} G. Gibbons, S. Hawking, Phys. Rev. D {\bf 15}, 2752 (1977). 
\bibitem{parikh} M. Parikh, F. Wilczek, Phys. Rev. Lett. {\bf 85}, 5042 (2000). 
\bibitem{Robwilczek} S. P. Robinson and F. Wilczek, Phys. Rev. Lett. {\bf 95},
 011303 (2005) [gr-qc/0502074].
\bibitem{Isowilczek} S. Iso, H. Umetsu and F.Wilczek, Phys. Rev. Lett.{\bf 96}, 151302 (2006) [hep-th/0602146].
\bibitem{shailesh2} R. Banerjee, S. Kulkarni, Phys. Lett. B {\bf 659}, 827 (2008), arXiv:0709.3916 [hep-th].   
\bibitem{shailesh} R. Banerjee, S. Kulkarni, Phys. Rev. D {\bf 77}, 024018 (2008) arXiv:0707.2449 [hep-th].
\bibitem{rabin} R. Banerjee, arXiv:0807.4637 [hep-th]. 
\bibitem{york} J.W.York,Jr., Phys. Rev. D 31, 755 (1985).
\bibitem{lousto} C.O.Lousto and N.Sanchez, Phys. Lett. B 212, 411 (1988).
\bibitem{fursaev} D.V.Fursaev, Phys. Rev. D 51, 5352 (1995) [arXiv:hep-th/9412161].
\bibitem{fabbri} R. Balbinot, S. Fagnocchi, A. Fabbri and G. Procppio, Phys. Rev. Lett. {\bf 94}, 161302 (2005), [gr-qc/0405096].
\bibitem{bibhas} R. Banerjee, B. Majhi, Phys. Lett. B {\bf 662}, 62 (2008), 
arXiv:0801.0200 [hep-th].
\bibitem{shailesh3} S. Gangopadhyay, S. Kulkarni, Phys. Rev. D {\bf 77}, 024038 (2008), arXiv:0710.0974 [hep-th].
\bibitem{sunandan} S. Gangopadhyay, Phys. Rev. D {\bf 77}, 064027 (2008), 
 arXiv:0712.3095 [hep-th].
\bibitem{peng} J. J. Peng and S. Q. Wu, arXiv:0801.0185 [hep-th].
\bibitem{Iso} S. Iso, T. Morita and H. Umetsu, arXiv:0710.0456 [hep-th].
\bibitem{bertlmann} R. Bertlmann, "Anomalies In Quantum Field Theory,"
(Oxford Sciences, Oxford, 2000). 
\bibitem{Fujikawa} K. Fujikawa and H. Suzuki, "Path Integrals and Quantum
 Anomalies, " (Oxford Sciences, Oxford, 2004).
\bibitem{Bardeen} W. A. Bardeen, B. Zumino, Nucl. Phys. B {\bf 244}, 421 (1984).
\bibitem{Isoumtwilczek} S. Iso, H. Umetsu and F. Wilczek, Phys. Rev. D {\bf 74}, 044017 (2006) [hep-th/0606018].
\bibitem{Leut} H. Leutwyler, Phys. Lett. B {\bf 153}, 65 (1985).
\bibitem{kohlprath} R. Bertlmann and E. Kohlprath, Ann. Phys. (N.Y) {\bf 288}, 137 (2001).
\bibitem{fabbri1} R. Balbinot, A. Fabbri, I. Shapiro, Nucl. Phys.B {\bf 559}, 301 (1999).
\bibitem{fabbri2} R. Balbinot, A. Fabbri, I. Shapiro, Phys. Rev. Lett. {\bf 83}, 1494 (1999).
\bibitem{mottola} P. R. Anderson, E. Mottola, R. Vaulin, Phys. Rev.D {\bf 76}, 124028 (2007) arXiv:0707.3751 [gr-qc]. 

\end{thebibliography}
\end{document}